# SLID-ICV Vertical Integration Technology for the ATLAS Pixel Upgrades

A. Macchiolo[a,*], L. Andricek,[a,b], H.G. Moser[a,b], R. Nisius[a], R.H. Richter[a,b], P. Weigell[a]

*a Max-Planck-Institut für Physik, Föhringer Ring 6, D-80805, Munich, Germany*
*b Max-Planck-Institut Halbleiterlabor, Otto-Hahn Ring 6, D-81739, Munich, Germany*

**Abstract**

We present the results of the characterization of pixel modules composed of 75 μm thick n-in-p sensors and ATLAS FE-I3 chips, interconnected with the SLID (Solid Liquid Inter-Diffusion) technology. This technique, developed at Fraunhofer-EMFT, is explored as an alternative to the bump-bonding process. These modules have been designed to demonstrate the feasibility of a very compact detector to be employed in the future ATLAS pixel upgrades, making use of vertical integration technologies. This module concept also envisages Inter-Chip-Vias (ICV) to extract the signals from the backside of the chips, thereby achieving a higher fraction of active area with respect to the present pixel module design. In the case of the demonstrator module, ICVs are etched over the original wire bonding pads of the FE-I3 chip. In the modules with ICVs the FE-I3 chips will be thinned down to 50 um. The status of the ICV preparation is presented.

## 1. Introduction

An intense R&D activity is being carried out to implement vertical integration technologies in the pixel modules for the future upgrades of the tracking detectors at LHC to achieve the maximal active area and a reduction of the material budget. One main ingredient of 3D technologies is represented by the Through-Silicon-Vias (TSV) to route the signals vertically through the read-out chip. These can be implemented following two different approaches: Via First and Via Last. The Via Last approach, where vias are etched after the front-end CMOS processing, applied to tracking detectors in high energy physics, allows for the extraction of signal and services across the chip to the backside. In a 3D compliant design of the pixel electronics, TSVs could eventually avoid the need for the cantilever area where the wire bonding pads are presently located.

In the context of vertical integration technologies new sensor-chip connection methods can also be explored. The SLID process was applied for mechanical and electrical interconnects in combination with the TSV technology of the Fraunhofer EMFT [1] and it was introduced as ICV (Inter-Chip-Via) technology as a fully modular concept for vertical system integration optimized for chip-to-wafer stacking [2,3,4].

\* *E-mail address*: annamac@mpp.mpg.de.

## 2. SLID Interconnection Technology

The SLID interconnection technology, representing an alternative to the standard bump-bonding, is characterized by a very thin eutectic Cu-Sn alloy. To prepare the sensor and the readout chip for SLID, a 100nm thin TiW diffusion barrier is placed on the metallized contact pads. This has been shown to prevent atoms from the 5 μm copper layers electroplated on both devices to diffuse into the silicon [5]. In addition, on one of the two copper layers a 3μm thin layer of tin is applied. To form the connection the two devices are aligned, brought in contact, and heated to a temperature of around $(240 - 320)°C$. At this temperature the tin diffuses into the copper to form the Cu-Sn alloy. As the melting point of this alloy is around 600°C multiple layers can be stacked and connected without melting the previous SLID connections.

## 3. FE-I3 module as demonstrator for the ICV technology

The work presented aims at a new detector concept in the framework of the ATLAS Inner Tracker upgrade. In the first part of the R&D project, demonstrator modules composed of pixel sensors, with an active thickness of 75 μm, connected to the ATLAS FE-I3 chips by SLID, have been realized. In this case the modules are read-out through a fan-out structure on the sensor. For this initial step, the FE-I3 chips have been thinned down to a thickness of 200 μm. In the second part of the project ICVs will be etched on the original wire bonding pads of the chips, and the modules will be read-out from the backside of the ASIC, thinned to 50 μm (Fig.1).

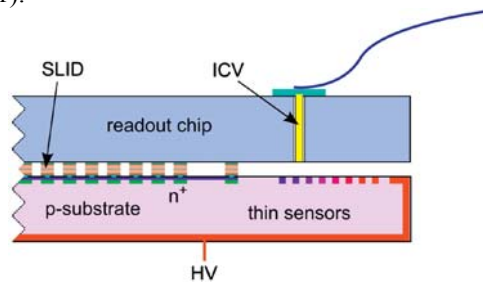

Fig. 1: Demonstrator module with thin sensors and ICV-SLID vertical integration technology. ICVs are etched on the chip pads originally designed for wire bonding.

The n-in-p pixel sensors to be interconnected by SLID have been produced from wafers of standard thickness using the MPP-HLL thinning process [6]. In total 8 n-in-p wafers, with active thickness of 75 µm and 150 µm, have been processed and characterized. The pre-irradiation measurements show a very good device yield and high breakdown voltages [7]. After irradiations of bare sensors up to a fluence of $10^{16}$ $n_{eq}$/cm$^2$ charge collection efficiency measurements yield charges close to pre-irradiation values for the 75 µm thick sensors. [8].

## 4. Interconnection and characterization of the SLID modules

The electroplating needed for the preparation of the SLID metal systems is applied to sensors and chips at the wafer level. The dimensions of the SLID pads on the chip and sensor sides are 27x60 µm$^2$, and they are designed in correspondence to the passivation openings originally foreseen for the bump bonding. The SLID interconnection is achieved through a chip to wafer technique where the chips are singularized and reconfigured onto a 6-inch handle wafer, according to the location of the pixel devices in the MPP-HLL sensor wafers.

The "pick and place" procedure of the chips in the handle wafer was affected by a misalignment with respect to the nominal positions, varying from 10 µm to 25 µm for the five structures where the misalignment did not cause shorted or open connections between the sensor and the chip pads. The positions of these five modules in the wafer are illustrated in Fig.2.

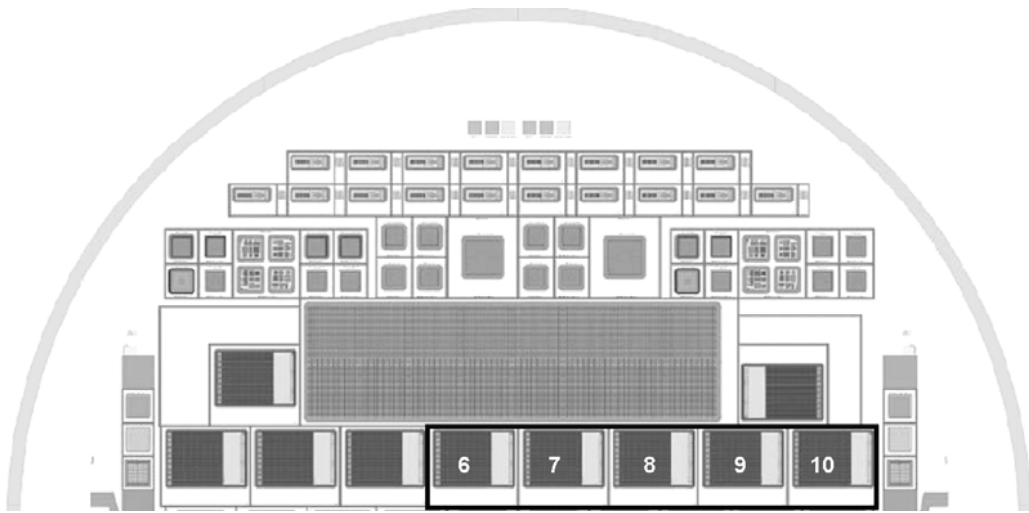

Fig. 2: Positions of the five modules (from SLID6 to SLID10) with a sufficient placement precision of the chips in the handle wafer not to cause shorted or open connections between the sensor and the chip pads.

The misalignment measured in the other five FE-I3 modules produced with this wafer resulted to be larger than the pixel pitch, inducing shorted connection between neighboring pixels. These structures were not further analyzed.

After the SLID interconnection the modules were diced and connected to a modified version of the ATLAS pixel detector board with a reverted order of the wire bonding pads. The new board reflects the arrangement of the wiring bonding pads on the SLID modules, differing from the standard FE-I3 detectors for an additional fan-out structure on the sensor side.

The SLID modules were tested with the USBpix system, developed within the ATLAS Pixel Group to read-out the FE-I3 assemblies [9]. The FE-I3 chip utilizes the time over threshold (ToT) technique to digitize the input charge. The threshold can be set via a 7-bit DAC (Threshold-DAC, TDAC) that allows for an adjustment for each pixel. The tuning procedure allows obtaining a homogenous threshold setting among different pixels, eliminating response differences, within the achievable threshold dispersion. Figure 3a shows the narrow threshold dispersion of 30e resulting after tuning the threshold of a SLID module to 2800e. Threshold dispersion and noise values (Fig.3b) are comparable to n-in-n and n-in-p FE-I3 modules interconnected with bump-bonding. [10].

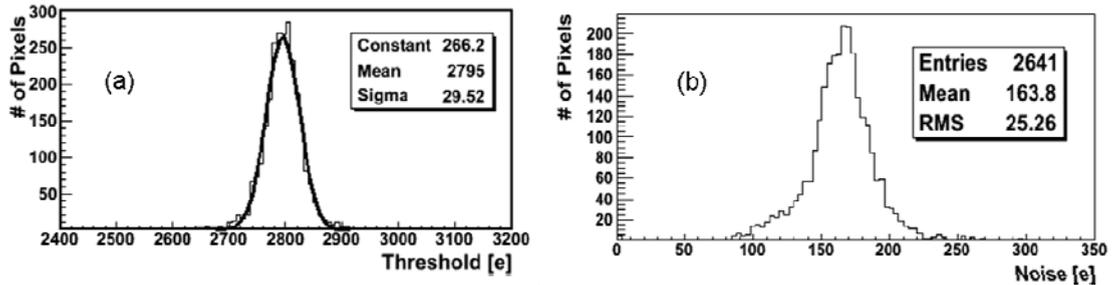

Fig. 3: (a) Threshold distribution and (b) noise distribution of one SLID interconnected module.

The reduced sensor active thickness of 75 μm that could in principle results in a higher noise due to the increased bulk capacitance, which is not observed.

Charge collection measurements have been performed using a $^{90}$Sr source, with an external trigger built by a scintillator and a photomultiplier, as shown in Fig.4a.

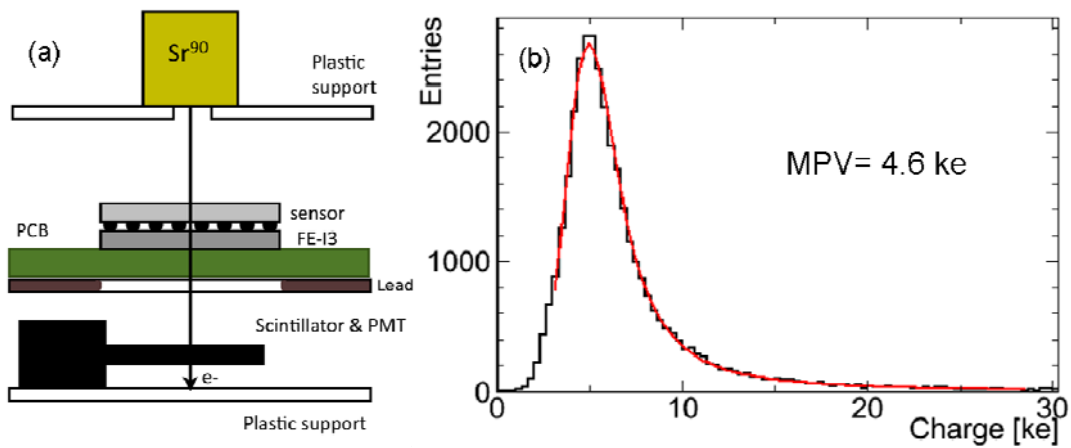

Fig. 4: a) Schematics of the set-up for the charge collection measurements with a radioactive source; b) Landau distribution of the charge collected with a $^{90}$Sr scan with a SLID module.

The collected charge measured for the SLID modules, varying among the 5 modules between 4.1 and 4.6 ke (Fig.4b), is in agreement with the charge measured in 285 μm thick detectors, connected to FE-I3 chips with bump-bonding, after scaling for the different active thicknesses [9]. The hit map for the module #10, is shown in Fig. 5.

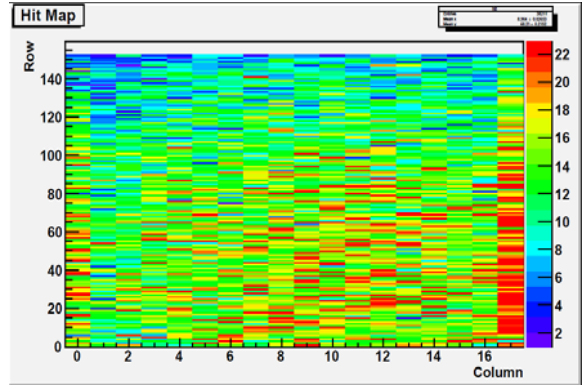

Fig. 5: Hit-map distribution for the SLID module #10, obtained with a $^{90}$Sr source scan. The number of hits per pixel shows that all channels are interconnected. The upper (white) rows were masked during data acquisition.

Not connected channels are identified by a different noise value in the threshold distributions and by the number of hits per pixel of the source scans. A higher number of not connected channels has been observed for the modules closer to the wafer centre (see Tab.1). The cause of the problem has been identified in the sensor passivation that is realized with Benzocyclobutene (BCB), used to achieve a good planarization of the surface needed for the electroplating and to improve the electrical isolation to the chip. The openings in the BCB that constitute the contacts to the underlying Aluminum layer suffer from a not complete etching and their dimensions decrease at smaller radii (Fig.6). This problem prevents in a fraction of the pixels the electrical contact between the Aluminum and the SLID pads. A possible solution for the additional sensor wafers still to be electroplated has been explored, consisting in a $SF_6$ plasma etching to enlarge the BCB openings. An optical inspection carried out after this treatment has shown an improvement of the contact definition over the full wafer.

Table 1. Number of disconnected channels in the five SLID modules that have been analyzed.

| Chip | Not connected channels | Percentage [%] |
|---|---|---|
| 6 | 731 | 30 |
| 7 | 713 | 29 |
| 8 | 274 | 11 |
| 9 | 134 | 6 |
| 10 | 0 | 0 |

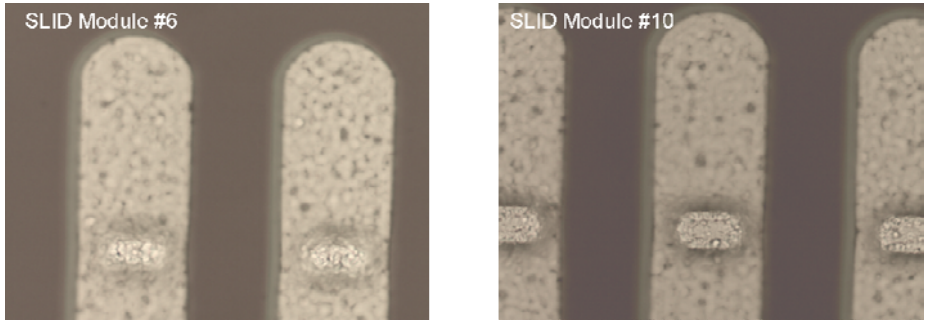

Fig. 6: Pictures of the contact openings in the BCB passivation layer for the SLID modules #6 and #10, taken in a sensor wafer not yet electroplated.

Furthermore the strength of the SLID connection has been estimated with pulling tests performed with the FE-I3 chips that were placed close to the electrically active modules to achieve a more homogenous pressure during the interconnection phase. The measured strength is of the order of 0.01N per connection, a value comparable to solder or indium bump-bonding.

## 5. Inter Chip Vias

The second phase of this R&D project foresees the extraction of the signals from the backside of the chip by using the ICV technology.
The planned process flow for the FE-I3 chips starts at wafer level with the etching of the vias on the chip pads originally designed for the wire bonding, after having removed the last aluminum layer. The vias cross-section has been optimized with etching trials in a FE-I3 wafer to be $3 \times 10$ $\mu m^2$, and the initial depth to 60 μm. This high aspect ratio is important to eventually transfer the process to a pixel-by-pixel level.

For isolation purposes a trench is etched around the multiple vias, with the same depth, as shown in Fig.7. For lateral via insulation a Chemical Vapour Deposition (CVD) of silicon dioxide is applied to ICVs and to the trench and then these structures are metalized with a tungsten filling.

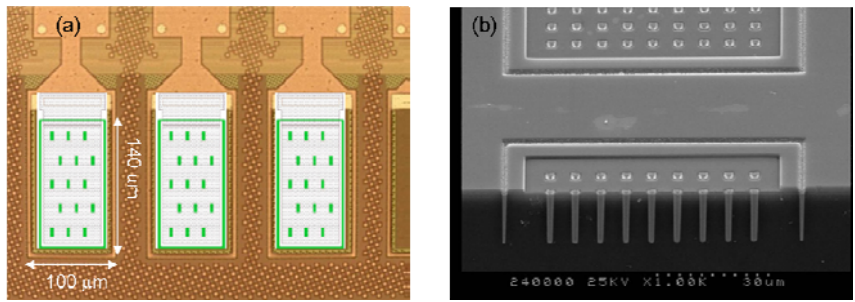

Fig. 7: (a) Wire bonding pads of the FE-I3 chip. The locations of the vias to be etched are shown in green, together with the shape of the isolation trench. (b) Picture showing ICVs enclosed in the isolation trench as realized in a different project at EMFT. The top side of the picture (light grey) corresponds to the surface of the wafer while the bottom part of the picture (dark grey) shows a cross section through the depth of the wafer.

After this step the electroplating is performed on the front side that is finally passivated. The chip wafer is then bonded to a handle wafer and thinned down from the backside to 50μm to expose the vias. An isolation layer is deposited on the backside and then opened in correspondence of the ICVs but not over the trench. Finally the metallization needed to create the new contact pads is applied. The subsequent SLID interconnection procedure is planned to follow the same steps taken in the first part of the project.

## 6. Conclusion

FE-I3 pixel modules interconnected with the SLID technology have been characterized before irradiation. For a fraction of the modules, a high number of not connected channels is due to not well defined contacts in the sensor passivation. The threshold and noise performance of all modules are very good and comparable to the standard n-in-n pixel technology with solder bump interconnection. Charge collection measurements performed with a $^{90}$Sr source yield MPVs of (4.1-4.6) ke, compatible with the pixel sensor active thickness of 75 μm. An optimization process of the ICV parameters has been completed. The R&D project will continue with the vias etching on the chips and the subsequent SLID interconnection of these chips to the pixel sensors.

### Acknowledgements

This work has been partially carried out within the RD50 Collaboration. We are grateful to the SiLAB group of the University of Bonn for the development and production of the detector board used in the measurements of the SLID modules.